\documentclass[aps,pre,twocolumn,groupedaddress,longbibliography]{revtex4-1}

\usepackage[english]{babel}

\usepackage[utf8]{inputenc}
\usepackage{times,mathptmx}
\usepackage{amsmath,amssymb}

\usepackage{graphicx} %

\usepackage{url}

\graphicspath{
	{./images/}
}

\usepackage{xcolor}
\usepackage{prettyref}
\newrefformat{fig}{Fig.\,\ref{#1}}
\newrefformat{sec}{sect.\,\ref{#1}}
\usepackage{siunitx}

\newcommand{\re}{\mathsf{Re}}

\newcommand{\cs}[1]{{#1}}
\begin{document}
\title{A flowing pair of particles in inertial microfluidics}
\author{Christian Schaaf}
\email{christian.schaaf@tu-berlin.de}
\author{Felix R\"uhle}
\author{Holger Stark}
\affiliation
{Institut f\"ur Theoretische Physik, Technische Universit\"at Berlin, Hardenbergstr. 36, 10623 Berlin, Germany}
\date{\today}
\begin{abstract}
A flowing pair of particles in inertial microfluidics 
gives important insights into
understanding and controlling the collective dynamics of particles like cells or droplets
in microfluidic devices. They are applied
in medical cell analysis and engineering. We study the dynamics of a pair of solid particles 
flowing through a rectangular microchannel 
using lattice Boltzmann simulations. We 
determine the inertial lift force profiles as a function of the two particle positions, their axial distance, and the Reynolds number. Generally, the profiles strongly differ between particles leading and lagging in flow and the lift forces are enhanced due to the presence of a second particle. At small axial distances, they
are 
determined by viscous forces, 
while inertial forces dominate at large separations.
Depending on the initial conditions, the two-particle lift forces in combination with the Poiseuille flow give rise to three types of \cs{unbound} particle trajectories, called moving-apart, passing, and swapping, and one type of \cs{bound} trajectories, where the particles perform damped oscillations. The damping rate scales with Reynolds number squared, since inertial forces are responsible for driving the particles to their steady-state positions.
\end{abstract}
\maketitle

\section{Introduction}

The control of hydrodynamic 
flow fields on a microscopic scale are required in a variety of different applications in  medicine, chemistry, and engineering \cite{martel_inertial_2014}. Microfluidic lab-on-a-chip devices allow to sample \cite{otto_realtime_2015} and sort cells \cite{hur_deformabilitybased_2011, hou_deformability_2010, schaaf_inertial_2017}, engineer flow patterns \cite{amini_engineering_2013}, and they can be used for 
fabricating metamaterials \cite{golosovsky_selfassembly_1999, kravets_engineering_2015}. While a lot of research 
has been and is still being done 
in the field of low-Reynolds-number flows \cite{batchelor_hydrodynamic_1972,darabaner_particle_1967,reddig_nonlinear_2013}, especially industrial applications need a high throughput in the microchannels \cite{zhang_fundamentals_2015}. 
The necessarily increased flow velocities 
initiated the field of inertial microfluidics. Here, fluid inertia is no longer 
negligible and new phenomena arise
\cite{dicarlo_inertial_2009}. 
One prominent example is the Segr\'e-Silberberg effect \cite{segre_radial_1961,segre_behaviour_1962},
where rigid particles assemble in an annulus, halfway between channel center and wall, when pumped through a cylindrical channel. 
Its first observation 
in 1961 inspired many experimental works \cite{bhagat_enhanced_2008,dicarlo_particle_2009} 
as well as analytical calculations \cite{asmolov_inertial_1999,hood_inertial_2015} 
and numerical simulations \cite{prohm_inertial_2012,chun_inertial_2006,kruger_interplay_2014,prohm_feedback_2014,yan_hydrodynamic_2007}. 
It can be rationalized by a lift-force profile, which a single particle experiences in the channel cross-section \cite{matas_lateral_2009,dicarlo_particle_2009, prohm_controlling_2014} 
\cs{and which can be used to implement an optimal-control scheme \cite{prohm_optimal_2013}.}

When the density of particles in 
the channel flow increases, they start to form microfluidic crystals or particle trains \cite{matas_trains_2004, lee_dynamic_2010}. 
Here,  the particles assemble 
in a linear or zig-zag pattern with a fixed axial distance 
typically ranging from 2.2 to
5 particle diameters \cite{humphry_axial_2010, kahkeshani_preferred_2016}. A deeper understanding of these particle trains is important for 
cell analysis \cite{edd_controlled_2008} and for understanding phonon 
excitations in microfluidic crystals \cite{beatus_phonons_2006,schiller_collective_2015}. As 
particle densities are 
still small, pair interactions of the particles can provide a first understanding.
At vanishing Reynolds numbers pair interactions 
were studied by \citet{batchelor_hydrodynamic_1972} in an unbounded shear flow.
They found open and closed trajectories for a pair of particles.
S\-imi\-lar trajectories also occur in Poiseuille flow \cite{reddig_nonlinear_2013}.
Now, including inertia has a profound influence. In particular, the
flow field around a single particle in a linear shear flow changes 
noticeably by losing 
the fore-aft symmetry
compared to low Reynolds numbers
\cite{subramanian_centrifugal_2006,subramanian_inertial_2006}. 
Applied to the trajectories of a particle pair,
\citet{kulkarni_pairsphere_2008} showed that 
closed trajectories in linear shear flow
are replaced by reversing and spiraling trajectories.
In microfluidic channels flowing particle pairs, when staying together, perform damped oscillations at finite Reynolds numbers \cite{lee_dynamic_2010,amini_inertial_2014}. This observation was one motivation for the work reported in this article.
In the following we study the dynamics of a pair of two solid particles 
driven by Poiseuille flow through
a rectangular microchannel. We 
perform lattice Boltzmann 
simulations of the Newtonian fluid and couple the 
particles to the fluid by the immersed boundary method. 
The
lift force profiles of both particles are crucially influenced by their neighbors and strongly depend on their distance along the
channel axis.
We find 
strong differences of the profiles 
for the particles leading and lagging 
in flow. Furthermore, lift forces in general are larger, 
which should enhance inertial focusing. 
Interestingly, how they scale with the Reynolds number depends on the axial particle distance. A
linear scaling at close distances reveals interactions determined by viscous forces while the
quadratic scaling for larger distances shows the dominating inertial forces.
Finally, we categorize
the different types of trajectories,
on which a particle pair moves, in terms of their initial lateral positions. When the particles stay together, damped oscillatory trajectories occur, which can be explained using the
two-particle lift force profiles.

The article is organized as follows. In Section 2 we explain the set-up of our system, describe our implementation of the lattice-Boltzmann method, and how we couple the particles to the fluid. In Section 3 we present the results for the lift force profiles and the particle trajectories. We summarize and close with final remarks in Section 4.

\section{Methods}

We study a pair of solid spherical particles moving in a microfluidic channel flow at moderate Reynolds numbers. In the following we shortly explain the microfluidic setup and 	the lattice-Boltzmann method to simulate the hydrodynamic flow.

\subsection{Microfluidic setup in the simulations}
\begin{figure}
	\includegraphics[width=.38\linewidth]{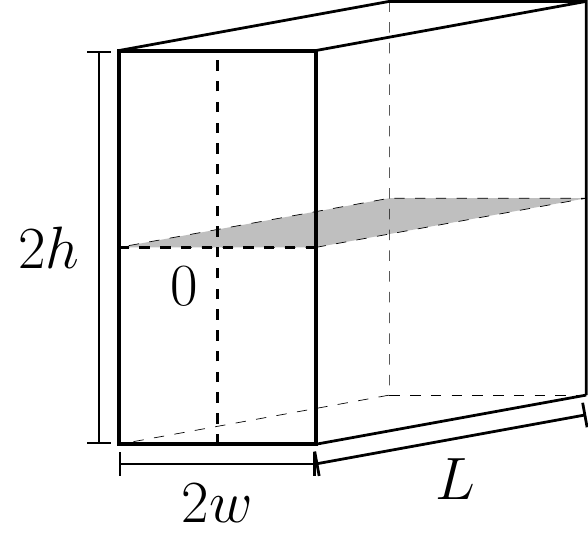}\hfil
	\includegraphics[width=.52\linewidth]{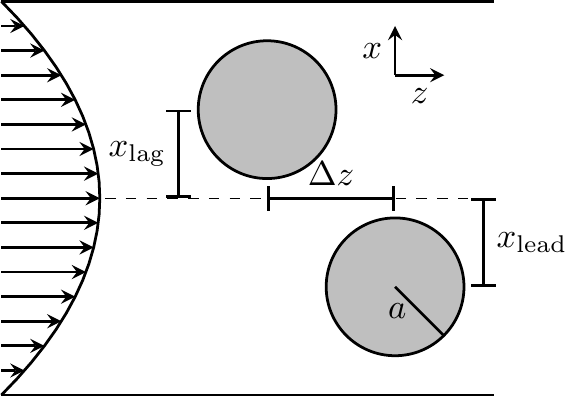}
	\caption{
		Left: A schematic of the microfluidic channel. We use a rec\-tangular channel with 
		width $2w$, height $2h$, and length $L$. 
		Only particle motion in 
		the $x,z$-plane (gray color) is considered. Right: Detailed view in the $x,z$-plane with Poiseuille
		flow 
		along the $z$-axis. 
		Two particles start with axial distance $\Delta z$ and lateral coordinates $x_{\mathrm{lead}}$, $x_{\mathrm{lag}}$ (measured against the center line) for particles leading and lagging in flow, respectively.
	}
	\label{fig:setup}
\end{figure}

The channel of length $L$ has a rectangular cross section (width $2w$ and height $2h$)
with an aspect ratio $w/h=0.5$
(see \prettyref{fig:setup}).
The channel is filled
with a Newtonian fluid with density $\rho$ and kinematic viscosity $\nu$. To drive 
a Poiseuille flow in the rectangular channel,
we apply a constant 
pressure force
\cite{bruus_theoretical_2008}. The flow is characterized by the channel Reynolds number $\re=2wu_\text{max}/\nu$ with the maximal flow velocity $u_\text{max}$. \cs{It is directed along the $z$-direction, which we call axial direction, while movements perpendicular to the flow direction  are referred to as lateral movements.}

In this Poiseuille flow we place two spherical particles with the same radius $a$ and neutral buoyancy.
At moderate Reynolds numbers fluid inertia becomes relevant and both particles experience lateral lift forces $f_\text{lift}$.
They push the particles into the $x,z$ plane containing the shorter cross-sectional axis \cite{prohm_feedback_2014}
and ultimately cause inertial focusing onto a specific position. To 
determine these lift forces in our simulations, we fix the particles' positions 
on the cross-sectional $x$ axis and measure the forces,
which the fluid exerts on them.
The lift forces are crucially influenced by the presence of the second particle and we will illustrate how they depend
on the axial particle distance $\Delta z$ in Sec.\ \ref{sec:equi_pos}. The particles flow with different velocity along the channel axis depending on their positions in the channel cross section.  So, when we measure the lift-force profiles, we let the particles move with their center-of-mass velocity and keep $\Delta z$ constant. This means that we effectively act with an axial force
\cs{along the flow direction}
on each particle, resulting in small changes of the lift forces according to the Saffman effect \cite{saffman_lift_1965}, as we demonstrate below.
In Sec.\ \ref{sec.parttra} we
analyze 
trajectories of the particle pair. Here, they can evolve freely without constraints.
Finally, along the flow direction we use periodic boundary conditions. To ensure that the particles do not interact with their mirror images, we use a channel length of $L=30a+\Delta z$.

\subsection{Lattice-Boltzmann method}
To solve the 
Navier-Stokes equations, we use the lattice Boltzmann method (LBM) in 3D 
based on 19 different velocities vectors (D3Q19) \cite{succi_lattice_2001} and rely on the Bhatnagar-Gross-Krook (BGK) collision
operator \cite{bhatnagar_model_1954}. In the LBM the fluid is modeled by a one-particle probability distribution $f_i(\vec x, t)$, which is 
determined on a cubic lattice with lattice spacing $\Delta x$. The distribution function depends on 
the lattice vectors $\vec x$ and the index $i$ stands for the 19 discretized velocity vectors $\vec c_i$ 
pointing to the edges and the faces of a cube and 
the zero velocity.
Now, $f_i(\vec x, t)$ evolves during time $\Delta t$ 	according to
two alternating steps:
\begin{align}
	\text{collision: }& f_i^\ast(\vec x,t)=f_i(\vec x,t)+\frac{1}{\tau}\left[f_i^\text{eq}(\vec x,t)-f_(\vec x,t)\right]\\
	\text{streaming: }& f_i(\vec x+\vec c_i\Delta t,t+\Delta t) = f_i^\ast(\vec x,t) \, ,
\end{align}
where $f_i^\text{eq}$ is a second-order expansion of the Maxwell-Boltzmann distribution  in the mean velocity and $\tau$ is the 
relaxation time 
of the BGK model. 

Macroscopic quantities like the density $\rho$ and the momentum density $\rho \vec u$ are defined via  the zeroth and first moments of the distribution function:
\begin{equation}
	\rho(\vec x_i,t)=\sum_i f_i(\vec x_i,t)
\end{equation}
\begin{equation}
	\rho(\vec x_i,t)\vec u(\vec x,t)=\sum_i \vec c_i f_i(\vec x_i,t)
\end{equation}

Typically, in the LBM the
density of the fluid is 
set to 1. 
The viscosity is related to the 
relaxation time $\tau$ \cite{dunweg_lattice_2008},
\begin{equation}
	\nu=c_s^2\Delta t\left(\tau-\frac 12\right) \, ,
\end{equation}
where
$c_s^2=1/3$ is the speed of sound measured
in LBM units. 
To ensure incompressibility of the fluid, simulations have to be performed at small Mach numbers $\mathrm{Ma}=u_\text{max}/c_s$. One additional constraint arises from the
immersed-boundary method, which we use 
to implement the two particles  (see Sec.\ \ref{subsec.particles}). It gives best accuracy for relaxation times $\tau \leq 1$ or  $\nu\leq 1/6$ \cite{kruger_efficient_2011}. In order to vary the Reynolds number, we fixed the viscosity to $\nu=1/6$ and modified the maximum flow velocity ensuring that $\mathrm{Ma}<0.1$, which corresponds to density 
variations of less than 1\%. 

The channel flow was driven by a constant body force 
according to the Guo-force scheme\cite{guo_discrete_2002} and we used regularized boundary conditions at the walls \cite{latt_straight_2008}. The lattice-Boltzmann simulations were performed with
the code provided by the Palabos project\cite{_palabos_2013},
which we supplemented by the implementation of particles using
an immersed boundary method. 
Finally, we discretized the width of the channel along the $x$ axis by 
75 lattice cells. 

\cs{
We use the same simulation code as in our previous publication \cite{prohm_feedback_2014}. We only added the 
event-based Euler method to prevent overlap between particles, which we describe at the end of the following section.}

\subsection{Modeling of solid particles}
\label{subsec.particles}

We modeled the colloids by an immersed boundary and couple them to the fluid by the method proposed by Inamuro \cite{inamuro_lattice_2012}. This  immersed boundary method ensures the no-slip boundary condition at the particle surface by iteratively refining the body force $g_i$ acting on the surrounding fluid nodes. 
To interpolate the fluid velocity at a position between the nodes, we follow \citet{peskin_immersed_2002} and use an interpolation kernel, which considers all 
neighboring nodes in a sphere with radius
$2\Delta x$. For further details on these methods we refer the reader to the original 
publications and our previous work\cite{prohm_feedback_2014}. 

Furthermore, the approach assumes that the particles are filled with a Newtonian fluid which is unphysical. In order to compensate for this, we also 
apply Feng's rigid body approximation \cite{feng_robust_2009} and add an additional
force 
acting on the particle so that it moves
like a solid particle. 
With all contributions the equations of motion for the colloids are given by
\begin{align}
	\vec r_i(t+\Delta t) &= \vec r_i(t) + \vec v_i(t)\\
	M\vec v_i(t+\Delta t) &= M\vec v_i(t)+\vec F_i^\text{fluid}+\vec F_i^\text{Feng}\\
	I\vec \omega_i(t+\Delta t) &= I\vec \omega_i(t)+\vec T_i^\text{fluid}+\vec T_i^\text{Feng} \, ,
\end{align}
where 
$i$ 
is the particle index, $\vec r$, $\vec v$, and $\vec \omega$ 
are, respectively, the position, velocity, and angular velocity. 
Finally, $M$ and $I$ 
stand for the mass and moment of inertia.

For some trajectories, which we show in Sec.\ \ref{sec.parttra}, the two particles touch each other. To avoid overlap, we use an event-based Euler step for the colloids. When the particles are so close that they would overlap in the next time step, we reduce the time step to $\Delta \tilde t$ so that
the particles just touch, 
perform the collision between the particles, and finish the remaining time step with length $\Delta t -\Delta \tilde t$ with the new values for the particle velocities and angular velocities. To realize the collision, we follow Ref.\ \cite{allen_computer_1987} and
consider two rough hard spheres \cite{allen_computer_1987} so that 
during collision also 
angular momentum is exchanged.

\section{Results} \label{sec:results}

\begin{figure}
	\includegraphics[width=0.98\linewidth]{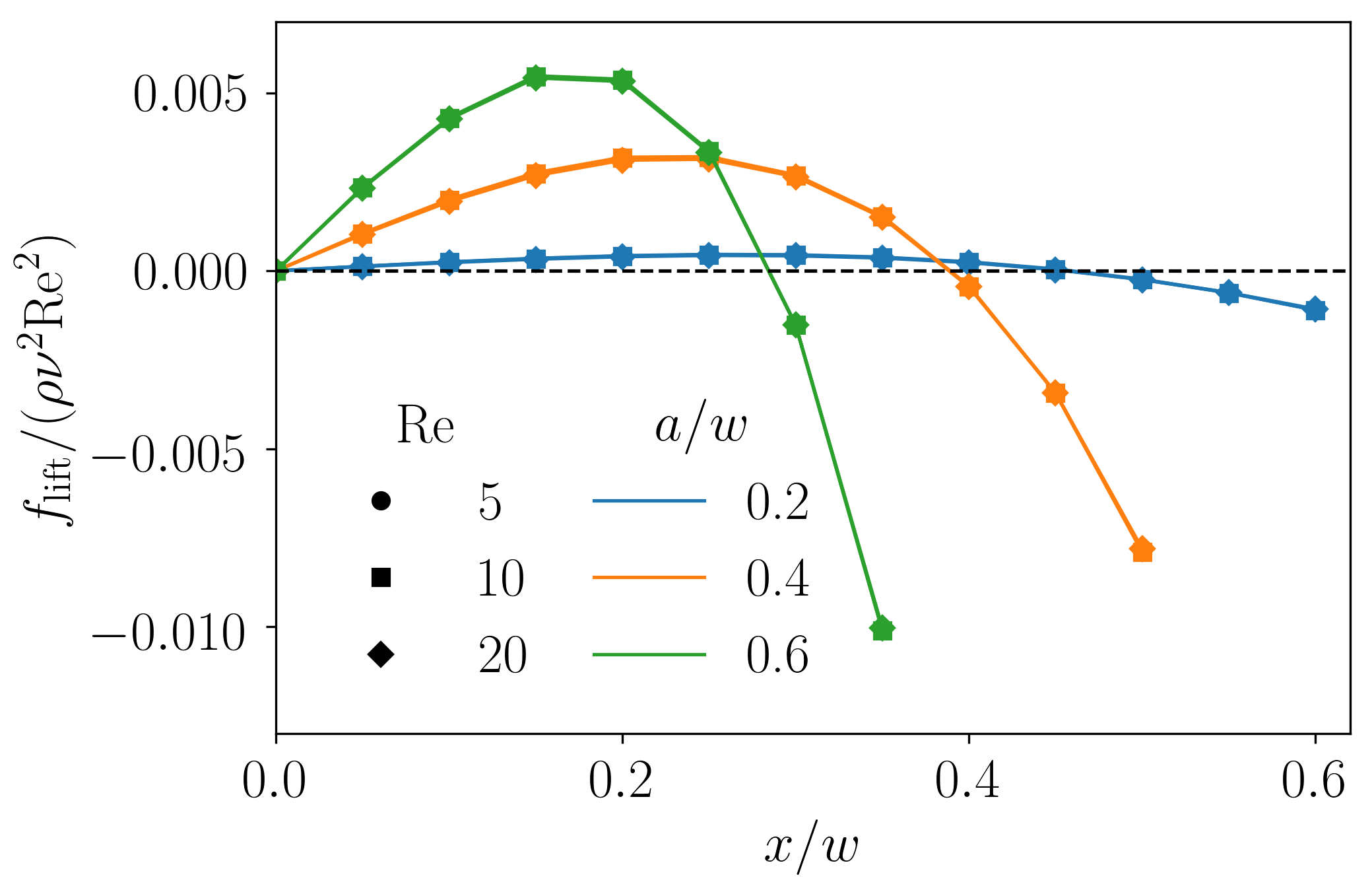}
	\caption{The lift force profile for a single particle depends on its radius $a$ and the channel Reynolds number $\re$. Scaled with $\re^2$ 
		the profiles for the same radius $a$ (the symbols represent data points) fall on top of each other and are hardly distinguishable.
	}
	\label{fig:force_profile_single}
\end{figure}

\subsection{Lift force profiles} \label{sec:equi_pos}

The dynamics of a particle pair in inertial microfluidics is best captured by the particle-particle lift force 
profiles.
They quantify the 
lift forces, the particles experience in the presence of the other particle either leading or lagging in flow [see Fig.\ \ref{fig:setup} for the geometry].
Zero forces correspond to 
fixed points or equilibrium positions in the channel cross section
and the magnitude of the force 
indicates how fast the particles 
become focused on their equilibrium positions. When both particles are mirrored at the channel axis the lift force reverses sign.
More importantly, when the flow direction is reversed such that the leading particle becomes lagging and vice versa, the lift force profiles change since due to secondary flow in the inertial regime the leading and lagging particles experience different flow fields. 

In previous work we already analyzed the lift-force profile for a single rigid particle \cite{prohm_inertial_2012,prohm_feedback_2014}. This force profile 
changes sign when mirrored at the channel center and scales with the Reynolds number $\propto\re^2$ (\prettyref{fig:force_profile_single}). Typically, one finds an unstable fixed point in the channel center and stable off-centered fixed points or equilibrium positions (indicated by a negative slope) along 
symmetry axes in the channel cross section. In this work we will focus on the two-particle lift-force profiles as a function of the axial particle distance $\Delta z$ \cs{(measured along the flow direction)}, the lateral coordinates $x_{\mathrm{lead}}$, $x_{\mathrm{lag}}$ and the channel Reynolds number $\re$ as indicated in \prettyref{fig:setup}. \cs{In what follows we concentrate on particles with radius $a/w=0.4$.}

\begin{figure}
	\includegraphics[width=\linewidth]{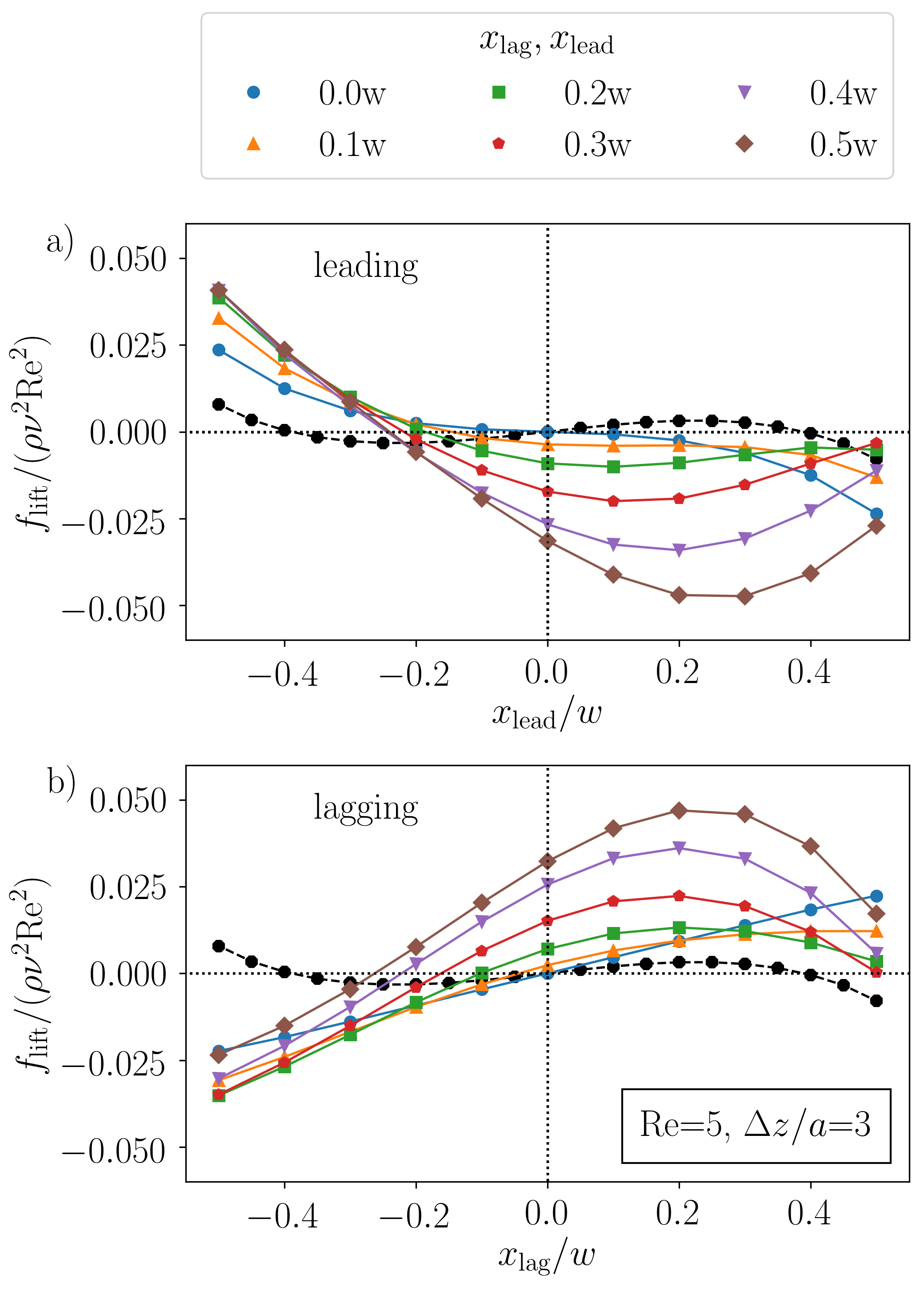}
	\caption{	Lateral lift force profiles along the short axis for the leading (a)	and lagging (b)	particle for $\re=5$,
		axial distance $\Delta z/a=3$, and particle radius $a/w=0.4$.
		The curve parameters are the positions of the lagging (a) or leading (b) particle, respectively.
		The black dotted line corresponds to the single-particle force profile. 
	}
	\label{fig:force_profile_lead_lag}
\end{figure}

\subsubsection{Parameter study}

In \prettyref{fig:force_profile_lead_lag} we demonstrate how the presence of another particle influences the lift force profiles and the equilibrium positions. We keep the axial distance of the particle pair fixed at $\Delta z = 3a$ and plot force profiles of the leading particle for different lateral positions $x_{\mathrm{lag}}$ of the lagging particle [ \prettyref{fig:force_profile_lead_lag}a)] and vice versa [\prettyref{fig:force_profile_lead_lag}b)]. Overall one recognizes that the profile is drastically influenced by an adjacent particle and lift forces generally are larger compared to the single-particle case. Thus, inertial focusing is enhanced.  

For the leading particle [see Fig.\ \ref{fig:force_profile_lead_lag}(a)]
we only find  stable fixed
points 
in the channel side opposite to the location of the lagging particle,  the other fixed points have disappeared. However, the new equilibrium positions
are closer to the channel center compared to the single-particle case (black line) and, ultimately, for $x_{\mathrm{lag}} = 0$ the stable fixed point is in the channel center.
In contrast, when the leading particle resides in the upper half of the channel, the fixed point of the lagging particle in the other channel side [\prettyref{fig:force_profile_lead_lag}b)] becomes unstable and stable equilibrium points only exist very close to the upper channel wall for sufficiently large $x_{\mathrm{lead}}$. Interestingly, the configuration with $x_{\mathrm{lead}} = x_{\mathrm{lag}} = 0$ is not stable. Finally, note that the lift-force profiles of the leading and lagging particles differ from each other due to secondary flow
as stated in the beginning. 

\begin{figure}
	\includegraphics[width=\linewidth]{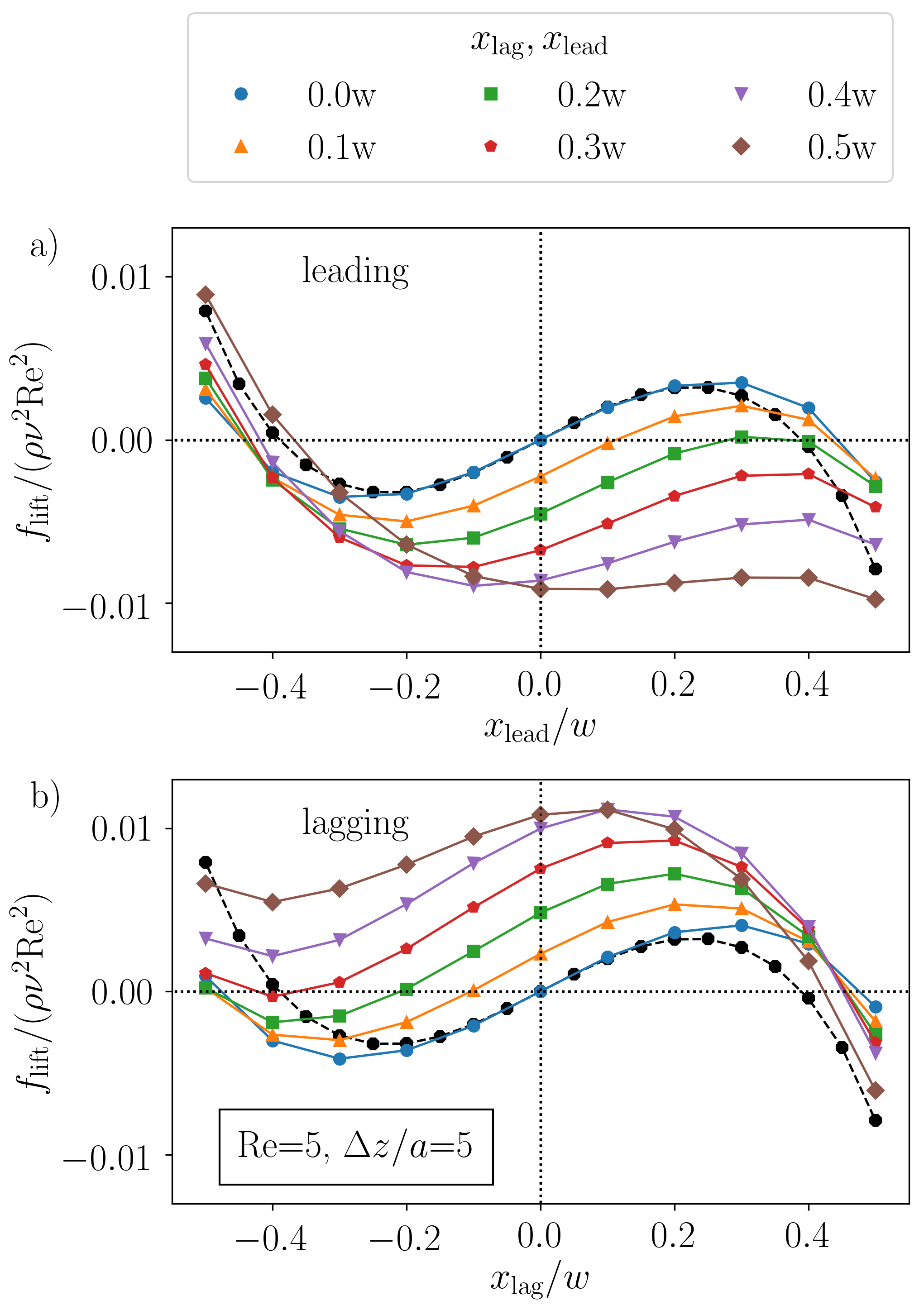}
	\caption{Lateral lift force profiles along the short axis for the leading (a) and lagging (b) particle for $\re=5$, particle radius $a/w=0.4$, and at the larger axial distance $\Delta z/a=5$ compared to \prettyref{fig:force_profile_lead_lag}. The curve parameters are the positions of the lagging (a) or leading (b) particle, respectively. The black dotted lines correspond to the single-particle force profile.
}
	\label{fig:force_profile_distance}
\end{figure}

\begin{figure}
	\includegraphics[width=\linewidth]{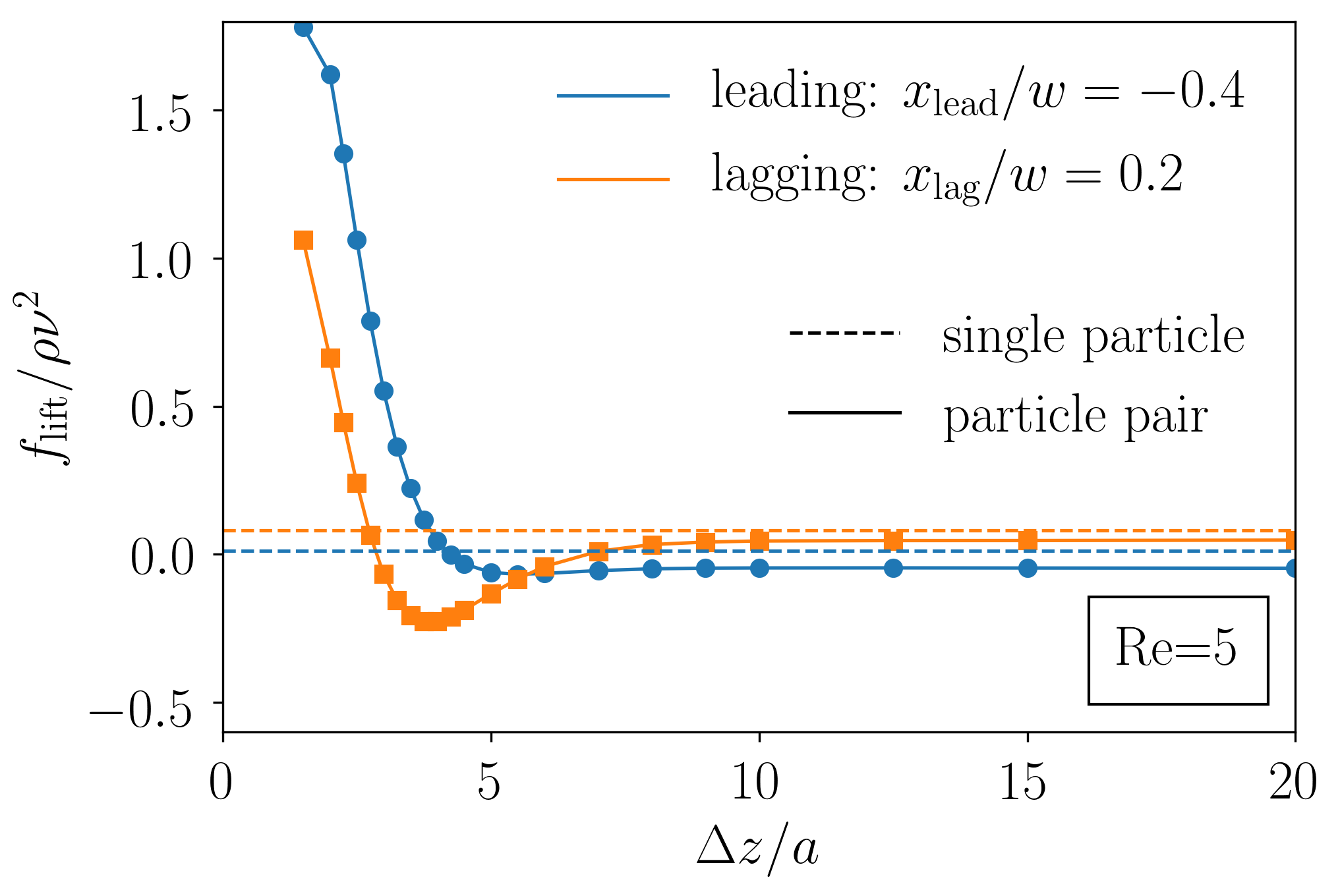}
	\caption{Lateral forces for a particle pair as a function of the axial distance $\Delta z$ with the 
		leading particle at $x_\text{lead}/w=-0.4$ and the lagging particle at	 $x_\text{lag}/w=0.2$. The dotted lines
		show the single particle lift forces at $x_\text{lead}$ and $x_\text{lag}$, respectively.}
	\label{fig:force_pair_distance}
\end{figure}

When we increase the distance $\Delta z$ of the two particles along the flow direction,
the lift-force profiles are more similar in shape to the profile of
a single particle, however shifted downwards (leading particle) or upwards (lagging particle) (\prettyref{fig:force_profile_distance}). 
In particular, for the cases where the 
other particle is relatively close the the channel center ($x/w<0.3$),
two stable equilibrium positions in the two respective sides of the channel are still present. In addition, when
the 
neighboring particle is in the channel center ($x/w=0$), the force profile agrees with %
the single-particle case close to the channel center (blue lines in \prettyref{fig:force_profile_distance}) but the stable equilibrium positions are located closer to the channel walls. Finally, by
increasing the axial distance between the particle pair, 
the strength of the lift forces decreases compared to Fig.\ \ref{fig:force_profile_lead_lag}.

The existence of 
stable fixed points in the lateral force profiles of both particles does not necessarily define a stable particle configuration,
since particles closer to the channel center move faster than particles near the channel walls. For a stable pair configuration the fixed points of both particles have to be at the same distance from the channel center.  From Fig.\ \ref{fig:force_profile_distance} we observe that this might be possible for $x_\text{lag}/w = - x_\text{lead}/w $ around 0.4, which we will indeed confirm
further below in Sec.\ \ref{sec.parttra}. In contrast,  when the particles are close together, for example at $\Delta z/a = 3$ as in Fig.\ \ref{fig:force_profile_lead_lag}, such a stable pair configuration is not possible.

In \prettyref{fig:force_pair_distance}
we fix the 
lateral positions of the leading and lagging particles and show 
how the lift forces change
with the 
axial distance $\Delta z$.
Below $\Delta z/a \approx 3$ the lift forces for both particles are positive so that they are pushed in positive $x$ direction towards the upper channel wall.
Thus the leading particle at $x_\text{lead}/w=-0.4$ moves closer to the channel center while the lagging particle at $x_\text{lag}/w=0.2$ approaches the upper wall. So, the distance of the particles grows and effectively they are repelled from each other.
For an axial distance $\Delta z/a > 3$ the 
lift force on the lagging particle at $x_\text{lag}/w=0.2$ becomes negative. Thus, both particles are pushed together, which can be described as an effective attraction. Increasing $\Delta z$ further initiates further sign changes.
Finally, for long distances both particles approach the lift forces 
of the respective single particle 
(dotted line). 
The small offset between the pair force and the single particle force can be explained by the Saffman effect. To measure the lift force, we have to fix the axial distance of these particles. In doing so we effectively accelerate the lagging particle and decelerate the leading particle. This additional axial force \cs{along the flow direction} leads to 
the small contribution in the lateral direction and explains the offset. 

\begin{figure}
	\includegraphics[width=\linewidth]{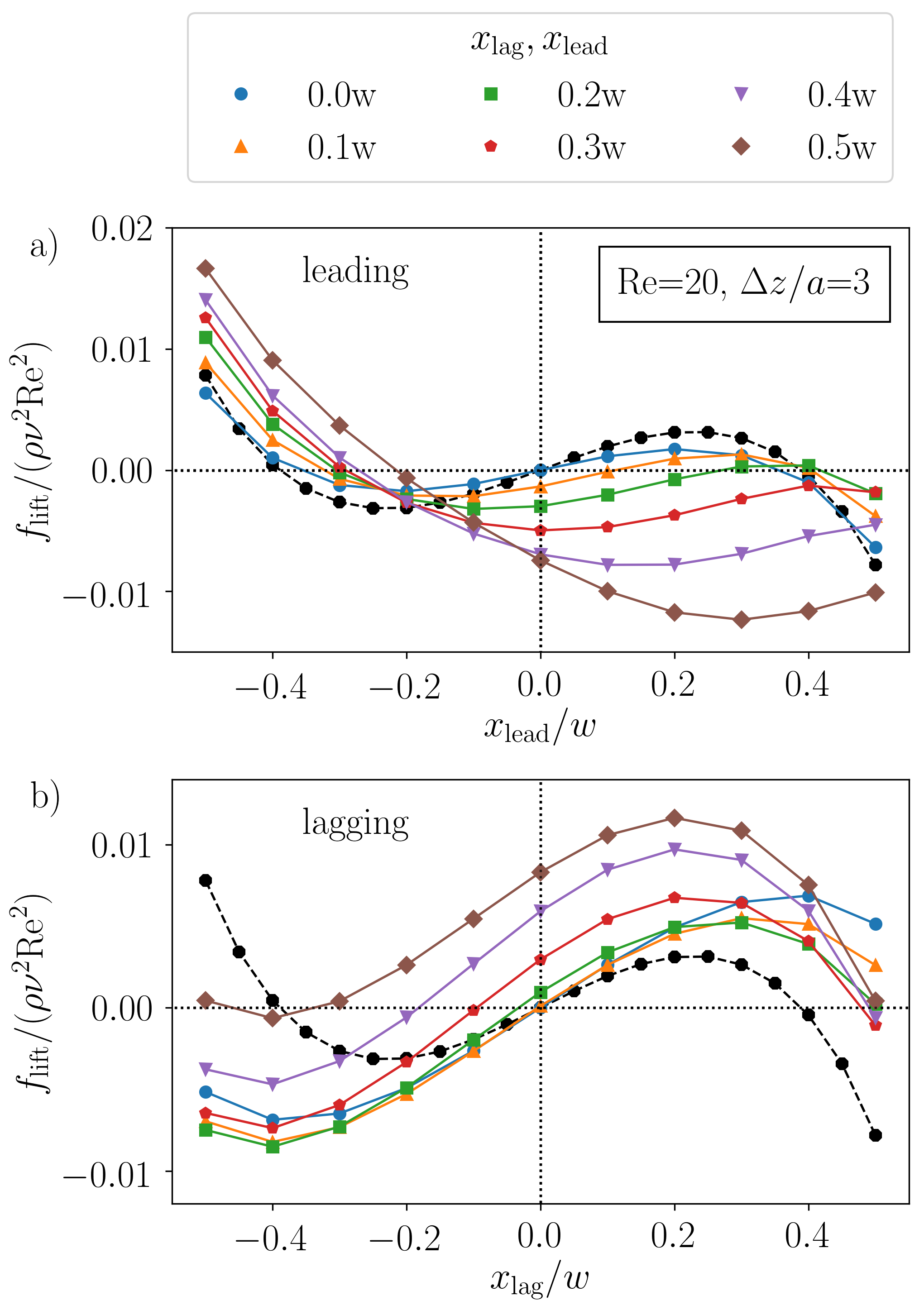}
	\caption{Lateral lift force profiles along the short axis for the leading (a)	and lagging (b) particle for $\re=20$, $\Delta z/a=3$, and $a/w=0.4$. The curve parameters are the positions of the lagging (a) or leading (b) particle, respectively. The black dotted lines correspond to the single-particle force profile.
	}
	\label{fig:force_profile_reynolds}
\end{figure}

Further, we
study how a variation of
the Reynolds number influences the two-particle force profiles. Again, we 
plot them at an axial distance 
$\Delta z/a=3$ as in \prettyref{fig:force_profile_lead_lag} but now for $\re = 20$ instead of $\re = 5$ (see \prettyref{fig:force_profile_reynolds}).
We immediately recognize that in contrast to \prettyref{fig:force_profile_lead_lag} the force profiles are similar in shape
to the one-particle profile but shifted 
upwards (lagging particle) or 
downwards (leading particle) with increasing lateral distance.
We saw a similar behavior already in \prettyref{fig:force_profile_distance} for $\re=5$ at larger axial distance $\Delta z/a=5$. 
In both cases the strength of the lift forces are similar to the single-particle forces, while in \prettyref{fig:force_profile_lead_lag} 
the two-particle induced forces are considerably larger than the inertial forces on a single particle. In addition, the lift forces in 
\prettyref{fig:force_profile_reynolds} rescaled by $\rho \nu^2 \re^2$ are smaller than in \prettyref{fig:force_profile_lead_lag}, 
which suggest that the usual scaling with $\re^2$ does not apply. We study this in more detail in the next paragraph.

\subsubsection{Scaling of the lift force with $\re$}

\begin{figure}
	\includegraphics[width=\linewidth]{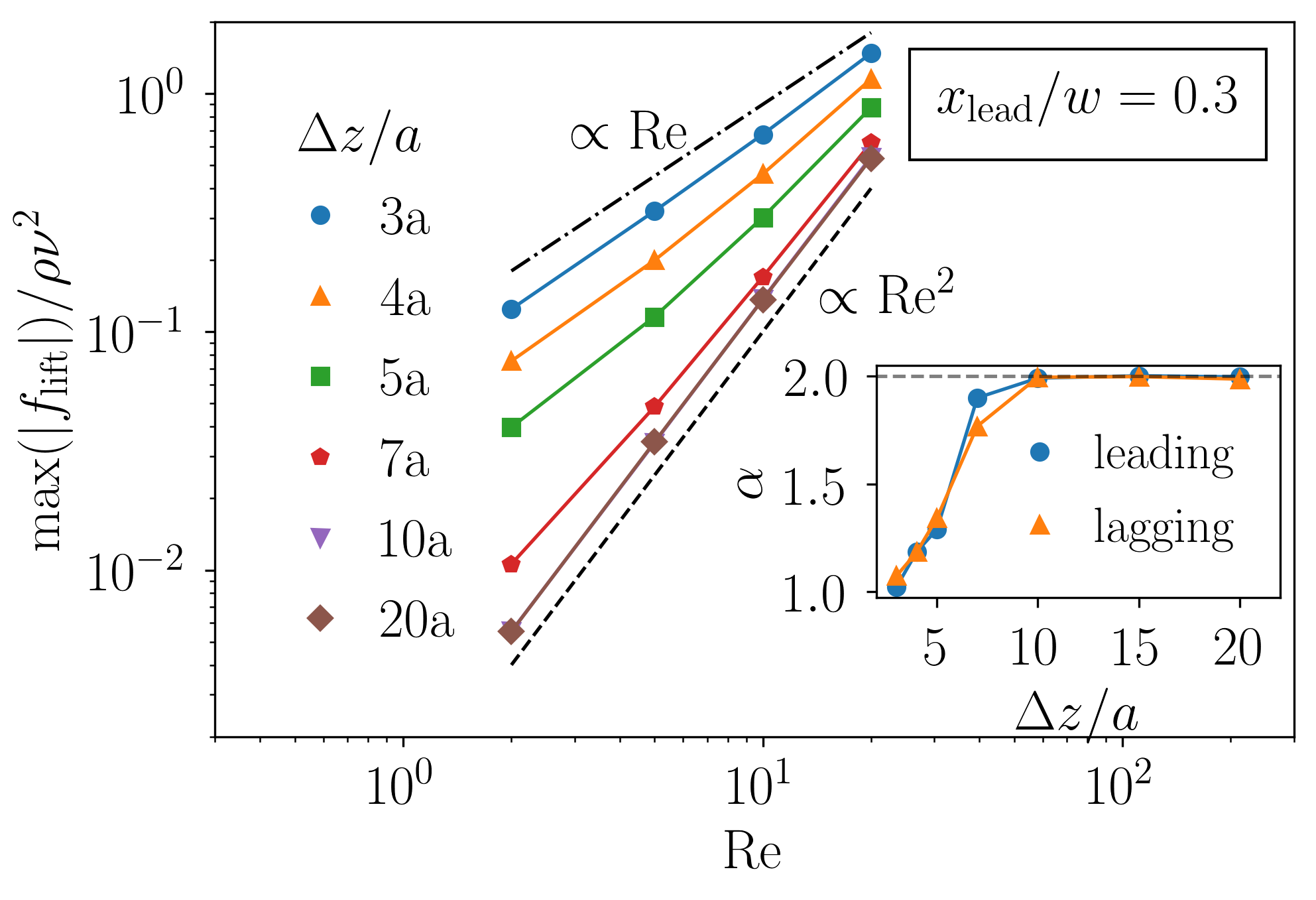}
	\caption{Maximum value of the lift force 
		of the lagging
		particle plotted versus $\re$ for different axial distances $\Delta z$. The leading  particle is fixed at 
		$x_\text{lead}/w=0.3$ while $x_\text{lag}$ is varied. The
		particle radius is $a/w = 0.4$. The dashed lines indicate scaling laws $\propto \re$ and $\propto \re^2$. Inset: Scaling exponent $\alpha$ from $f_\text{lift}^\text{max}\propto \re ^\alpha$ plotted versus $\Delta z$ for both particles. 
		Always the leading particle is fixed at
		$x_\text{lead}/w = 0.3$, while the position of the lagging particle is varied.
	}
	\label{fig:lift_force_scaling}
\end{figure}

We now take a closer look on 
how the lift force scales with the Reynolds number $\re$. We 
already realized that for small particle distances the two-particle lift forces no longer scale with $\re^2$ as in the single-particle case.
However, it is also clear that for large distances this scaling has to be 
recovered since the influence of the two particles on each other strongly decreases. To analyze this aspect in more detail, we 
fix the leading particle at $x_\text{lead}/w=0.3$ and vary $x_\text{lag}$.
We determine the maximum value of the magnitude of the 
lift force profile
for the 
lagging particle and plot it versus
Reynolds number for several particle distances
\footnote{In concreto, we
	consider the maximum value of the magnitude of the force within $|x|/w<0.4$ to ignore wall effects.}. 
Figure\ \ref{fig:lift_force_scaling} shows the results in double-logarithmic scale. One clearly recognizes a power-law scaling with exponent $\alpha$:
$f_\text{lift}^\text{max}\propto \re ^\alpha$. 
In the inset, we plot $\alpha$ versus $\Delta z$
for both
the leading and lagging particle.
Indeed, we
find $\alpha = 2$
for $\Delta z/a>7$. When the particles 
approach each other, the scaling
exponent decreases
to almost $\alpha=1$ for $\Delta z = 3a$. This scaling helps to further 
understand the character of the lift force, in particular, when two particles interact. A particle disturbs the fluid flow, which then influences the motion of nearby particles through a viscous coupling. This is the dominant contribution to the lift force at small distances as indicated by the linear scaling of the lift force with $\re$. The inertial contribution takes over at large distances, where the disturbance flow from the neighboring particle is weak, and one recovers the typical scaling for the inertial lift force, $f_\text{lift} \propto \re^2$. So, our analysis confirms the picture of Ref.\ \cite{lee_dynamic_2010}, which explicitly speaks about a viscous disturbance flow.

\cs{So, both Figs.\ \ref{fig:force_pair_distance} and \ref{fig:lift_force_scaling} indicate that beyond the distance $\Delta z/a \approx 7$ 
the particles essentially do not interact.  In Ref.\ \cite{haim_hydrodynamic_2009} 
it is argued that hydrodynamic interactions in a microchannel are screened on distances larger than the width of the channel cross section. In our case taking a particle radius of $a=0.4w$, a distance of $7a$ corresponds to $2.8w$, which is close to the channel width of $2w$. This explains our observation.}

\subsubsection{Contour plots}

\begin{figure}[bt]
	\includegraphics[width=\linewidth]{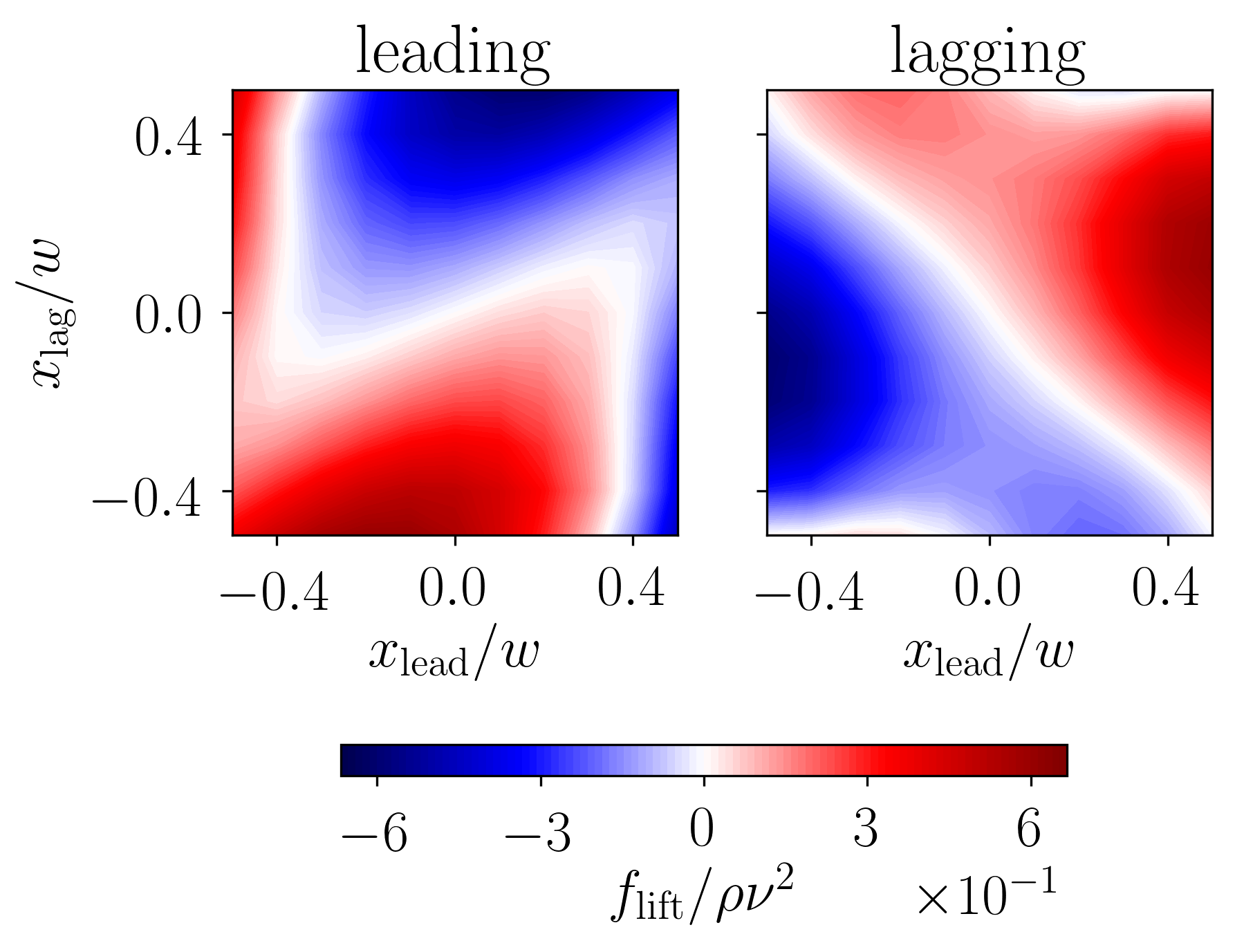}
	\caption{Color-coded lift force profiles in a two-dimensional representation plotted versus $x_\text{lead}$ and $x_\text{lag}$
		for the leading (left) and lagging (right) particle for 
		$\re=5$ and $\Delta z/a=4$. 
	}
	\label{fig:force_contour}
\end{figure}

In Sec.\ \ref{sec.parttra} we analyze 
possible trajectories for a pair of solid particles moving under the influence of the lateral lift forces. To
rationalize these trajectories,
it is instructive to use a two-dimensional representation of the respective lift-force profiles of the leading and lagging particles (see \prettyref{fig:force_contour}). Again, we clearly 
recognize the asymmetry of the profiles between the leading and the lagging particle, while each profile is symmetric under reflection at the
channel center. The white lines indicate zero crossings of the lift force, 
so stable and unstable equilibrium points.

\subsection{Two-particle trajectories} \label{sec.parttra}

\begin{figure}%
	\includegraphics[width=0.9\linewidth]{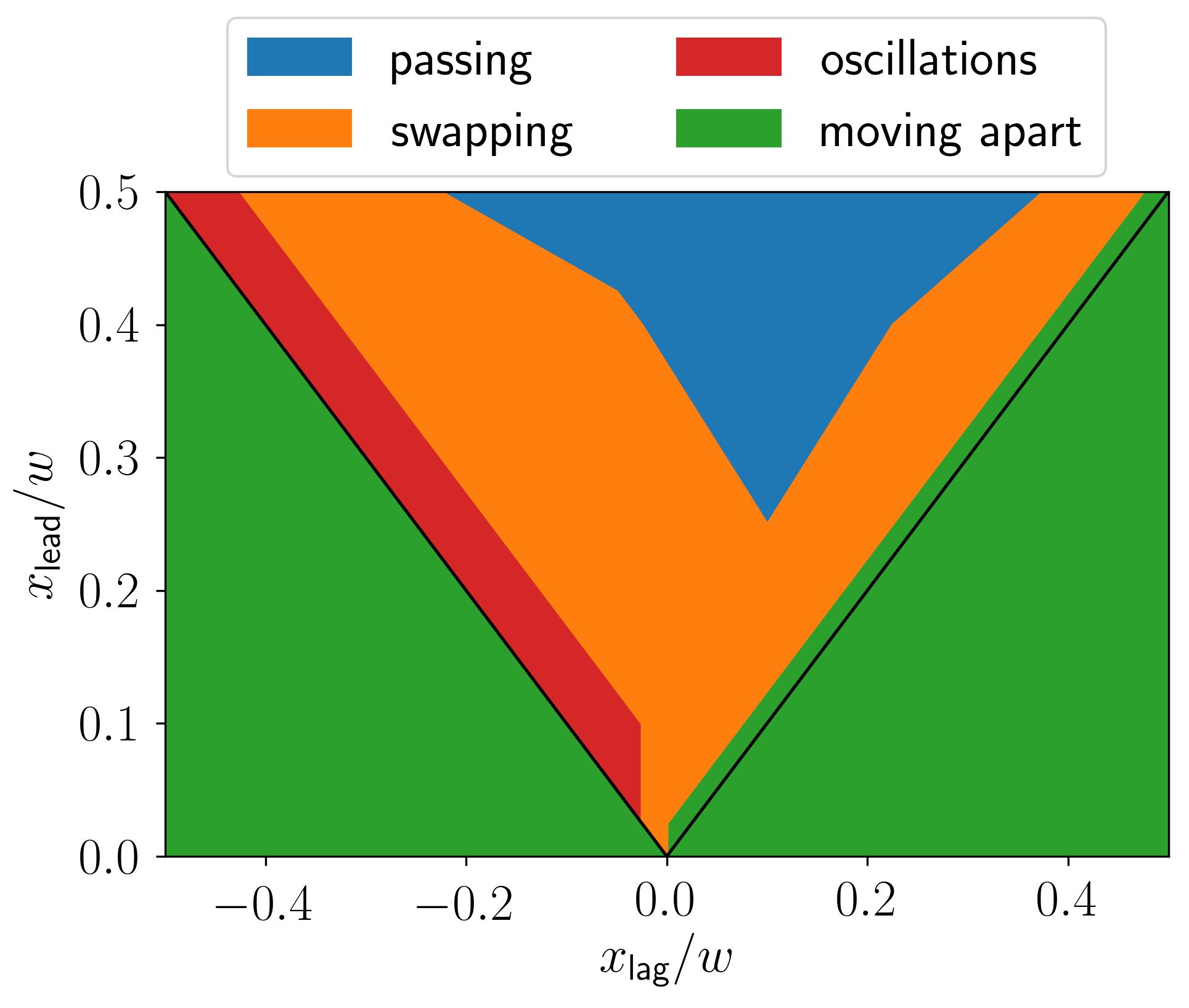}
	\caption{Types of particle trajectories indicated in parameter space of starting lateral positions $x_{\mathrm{lag}}$ and $x_{\mathrm{lead}}$
		for a pair of solid particles at $\re=10$. The starting axial distance is
		$\Delta z_0/a=5$ and particle radii are $a/w = 0.4$.
		The black lines indicate $|x_{\mathrm{lag}}| = |x_{\mathrm{lead}}|$.
	}
	\label{fig:trajectories}
\end{figure}

We now present the possible trajectories, which two particles traverse as a result of the coupled lift-force profiles presented 
in Sec.\ \ref{sec:equi_pos} and advection in the Poiseuille flow.
The different types occur depending on the starting lateral positions and the axial distance. Thus, in \prettyref{fig:trajectories} we categorize them in a diagram for the starting lateral positions $x_{\mathrm{lag}}$ and $x_{\mathrm{lead}}$, while keeping the 
starting axial distance fixed. We identified
four different kinds of coupled particle movements, which we term
moving apart, passing, swapping, and damped oscillations. 

\cs{The first three types of trajectories we name unbound as their particles}
drift apart and reach
their equilibrium lateral positions at large axial distances, where they do not influence each other anymore. 
We will %
\cs{ analyze these trajectories in more detail further below.}
We also observe bound trajectories, where the two particles ultimately perform damped oscillations about their equilibrium lateral positions. They occur in the narrow red region in \prettyref{fig:trajectories}, where the particles occupy opposing channel sides with the lagging particle only little faster than the leading such that they can stay together. 
\cs{We start with describing the bound trajectories.}
\subsubsection{Damped oscillations}
\begin{figure}[bt]
	\includegraphics[width=\linewidth]{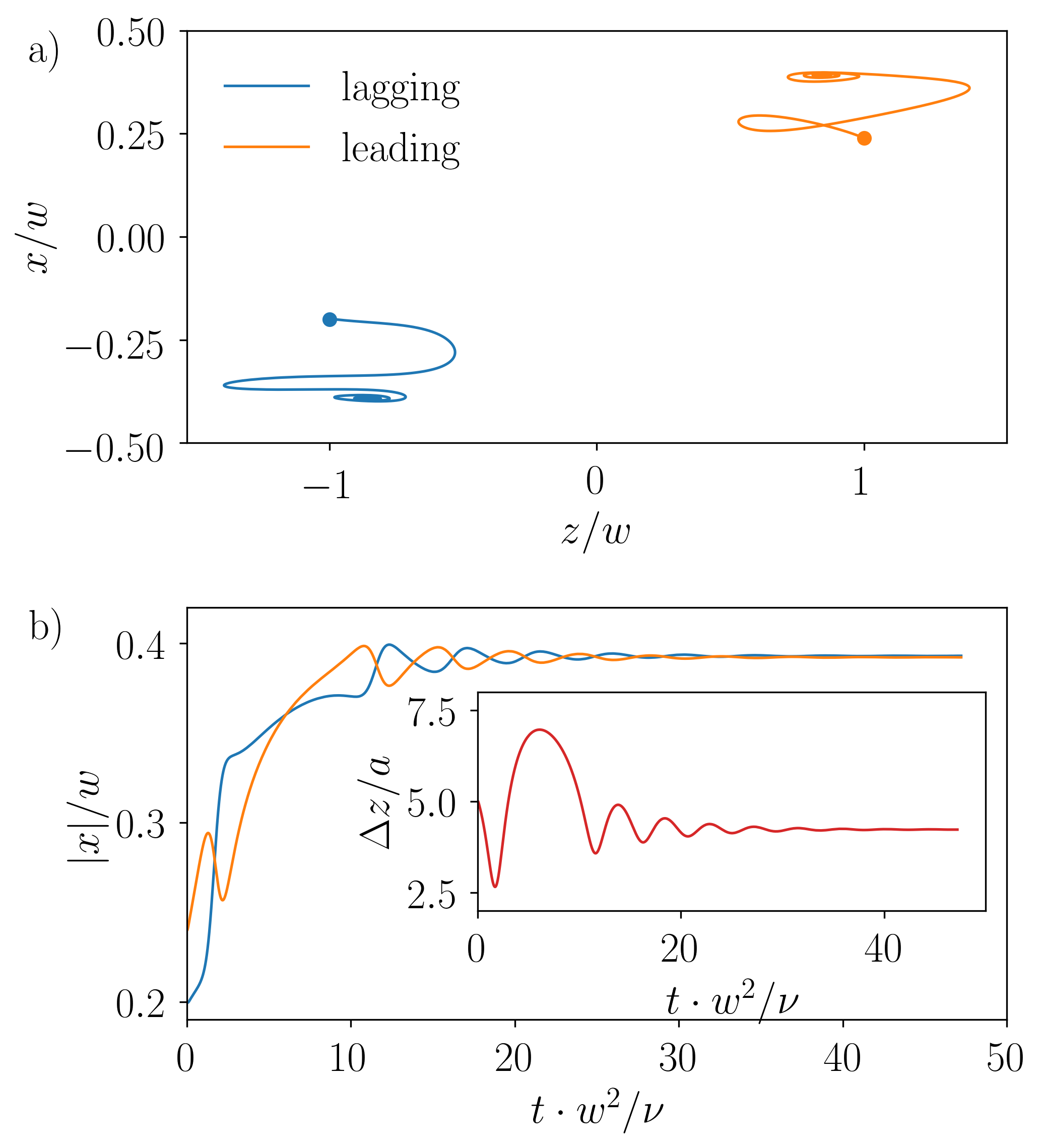}
	\caption{a)
		Trajectories of both particles in the $x,z$ plane drawn in the center-of-mass frame. The initial position is indicated by a dot. b) %
		Distance $|x|$ 	of each particle to the channel center plotted versus time. Inset: Axial distance $\Delta z$ of the particles versus time.
		The particles start with
		initial conditions $x_\text{lag}/w=-0.2$, $x_\text{lead}/w=0.24$, and $\Delta z_0/a=5$ and the Reynolds number is $\re=10$. 
	}
	\label{fig:oscillation_trajectory}
\end{figure}

Figure\ \ref{fig:oscillation_trajectory}(a) illustrates the damped oscillatory trajectories in the center-of-mass frame.
After a short transient 
regime at the beginning both particles migrate towards their 
stationary lateral positions ($|x_\text{eq}| / w\approx 0.4$), while performing damped oscillations with a strong difference
in the time-varying amplitudes along the channel axis and perpendicular to it 
[see 
Figs.\ \ref{fig:oscillation_trajectory}(a),(b) and inset].
In contrast to 
oscillatory trajectories also observed in pure
Stokes flow at small Reynolds number \cite{reddig_nonlinear_2013}, here the amplitudes decrease
in time  indicating that 
damping is an inertial effect. Such a 
damped motion is not possible in Stokes flow as it would violate 
kinetic reversibility of the Stokes equations. Interestingly, the damped oscillatory two-particle trajectories were also
observed in experiments by \citet{lee_dynamic_2010}. However, 
the authors report that the particles move apart symmetrically (``symmetric repulsive interactions''), while there is an 
asymmetry when approaching each other (``asymmetric attractive interactions''),
which we do not 
observe [compare \prettyref{fig:oscillation_trajectory}~b)]. Maximum and minimum displacements are always in phase. Ultimately, the particles reach their final lateral equilibrium positions, which agree with the positions of single particles.

\begin{figure}%
	\includegraphics[width=0.98\linewidth]{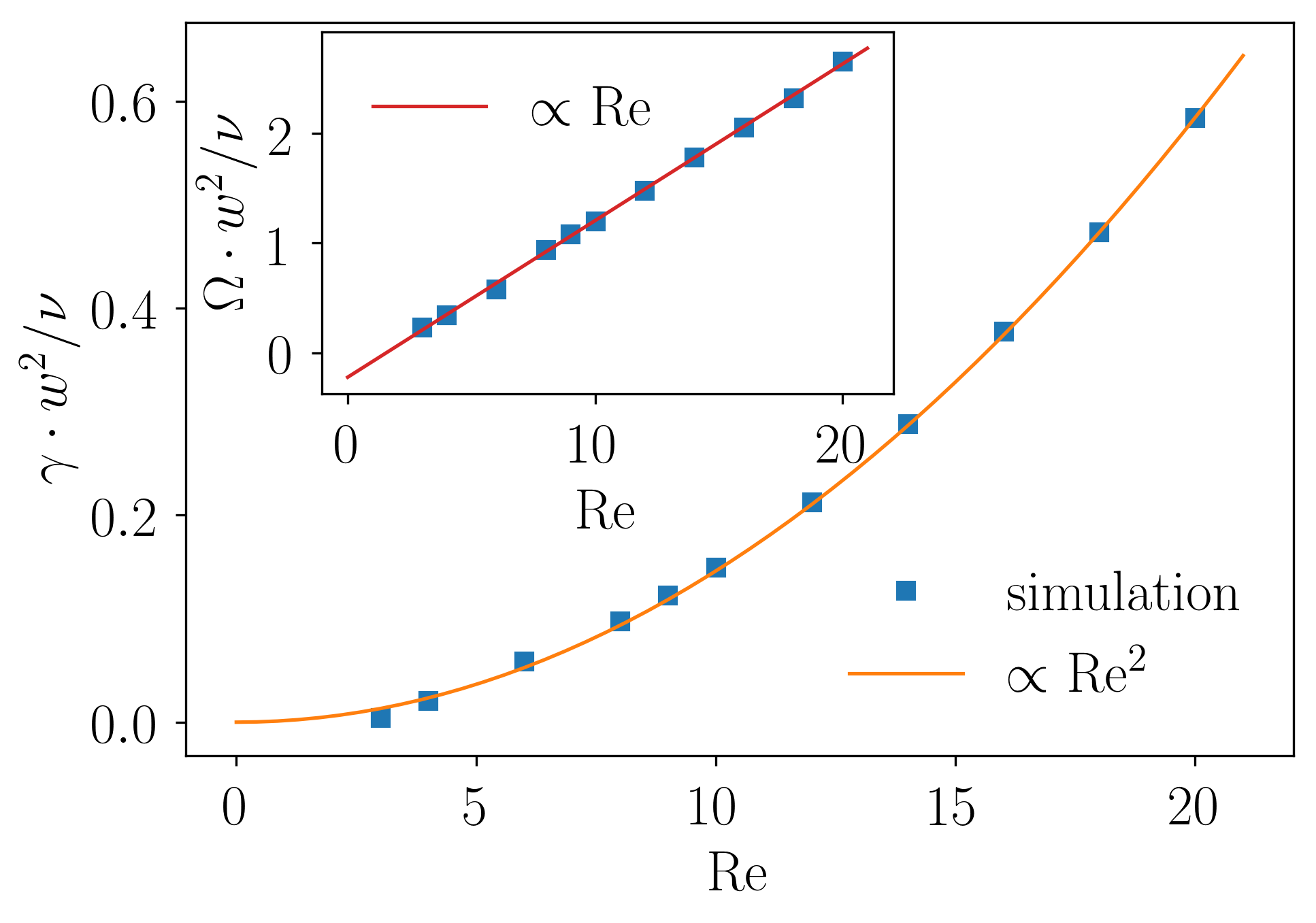}
	\caption{
		Damping rate $\gamma$ of the particle-particle distance and oscillation
		frequency $\Omega$ (inset) 
		plotted versus $\re$.
		The
		initial conditions are $x_\text{lag}/w=-0.2$, $x_\text{lead}/w=0.24$, and $\Delta z_0/a=5$. 
		Linear and quadratic fits in $\re$ are indicated, respectively.
	}
	\label{fig:oscillation_frequency}
\end{figure}

In the following we analyze how
oscillation frequency $\Omega$ and
damping rate $\gamma$
behave as a function of the Reynolds number (see \prettyref{fig:oscillation_frequency}). We determined 
$\Omega$ by measuring the time between maximal displacements and $\gamma$ by an exponential fit for the amplitudes decaying in time.
For the oscillation frequency $\Omega$ (inset of \prettyref{fig:oscillation_frequency}) we find a linear scaling with the Reynolds number,
which indicates that the oscillations 
are
due to the viscous coupling between the particles. In contrast, 
the damping 
rate scales quadratically with the Reynolds number since inertial lift forces drive them to their equilibrium positions. According to Fig.\ \ref{fig:lift_force_scaling} these inertial forces act here as a pertubation. Note that our findings on the damped oscillations are in full agreement with Ref.\ \cite{hood_pairwise_2017}.

The dynamics of the oscillating particle pair, which we discussed in Fig.\ \ref{fig:oscillation_trajectory}, can be nicely illustrated 
using
lift-force contour plots similar to the one we determined in 
Sec.\ \ref{sec:equi_pos} (see \prettyref{fig:force_contour} and video in SI) but now for $\re = 10$.
We start with the 
initial conditions
\[
x_\text{lag}/w=-0.2, \enspace x_\text{lead}/w=0.24,\enspace \text{and} \enspace \Delta z/a=5 \, .
\]
We can now follow the particles in the lift-force contour plots in video to understand their trajectories in the $x,z$ plane. In the beginning the lagging particle is faster as it is closer to the center. The signs of both 
lift forces are such that they push the particles 
towards the walls.
In this phase, since the lagging particle is faster the axial distance decreases. The leading particle turns around and moves	away from the upper wall, while the lagging particle still moves towards the lower wall. Thus, both forces are negative. 	Ultimately, the lagging particle is closer to the wall than the leading particle. It moves slower and the axial distance increases.
By following the trajectories further, the signs of the lift forces always indicate the lateral direction of the moving particles. 	In the end they show damped oscillations about the zero lines of the contour plots in agreement with the spiraling motion in the $x,z$ plane. Finally, after a few oscillations the particles reach their stable equilibrium positions, where the lift forces are zero.

Interestingly, we find that all 
bound particle pairs performing damped oscillations assemble at an axial distance of $\Delta z/a\approx 4.1$ independent of their initial conditions or the Reynolds number,
\cs{which we varied between 2 and 20.}
The value of 
\cs{this} 
axial distance is in good agreement 
with experimental \cs{ and theoretical} results \cite{humphry_axial_2010, kahkeshani_preferred_2016,matas_trains_2004, hood_pairwise_2017}. \cs{The scaling of the lift force with $\re$ (cf. \prettyref{fig:lift_force_scaling}) indicates that 
at this equilibrium distance the particle interactions are dominated by 
a viscous disturbance flow as already mentioned above and in Refs.\ \cite{lee_dynamic_2010,hood_pairwise_2017}. The shape of this flow does not depend on the Reynolds number, which explains
why the equilibrium distance is independent of $\re$.}

\begin{figure}%
	\includegraphics[width=\linewidth]{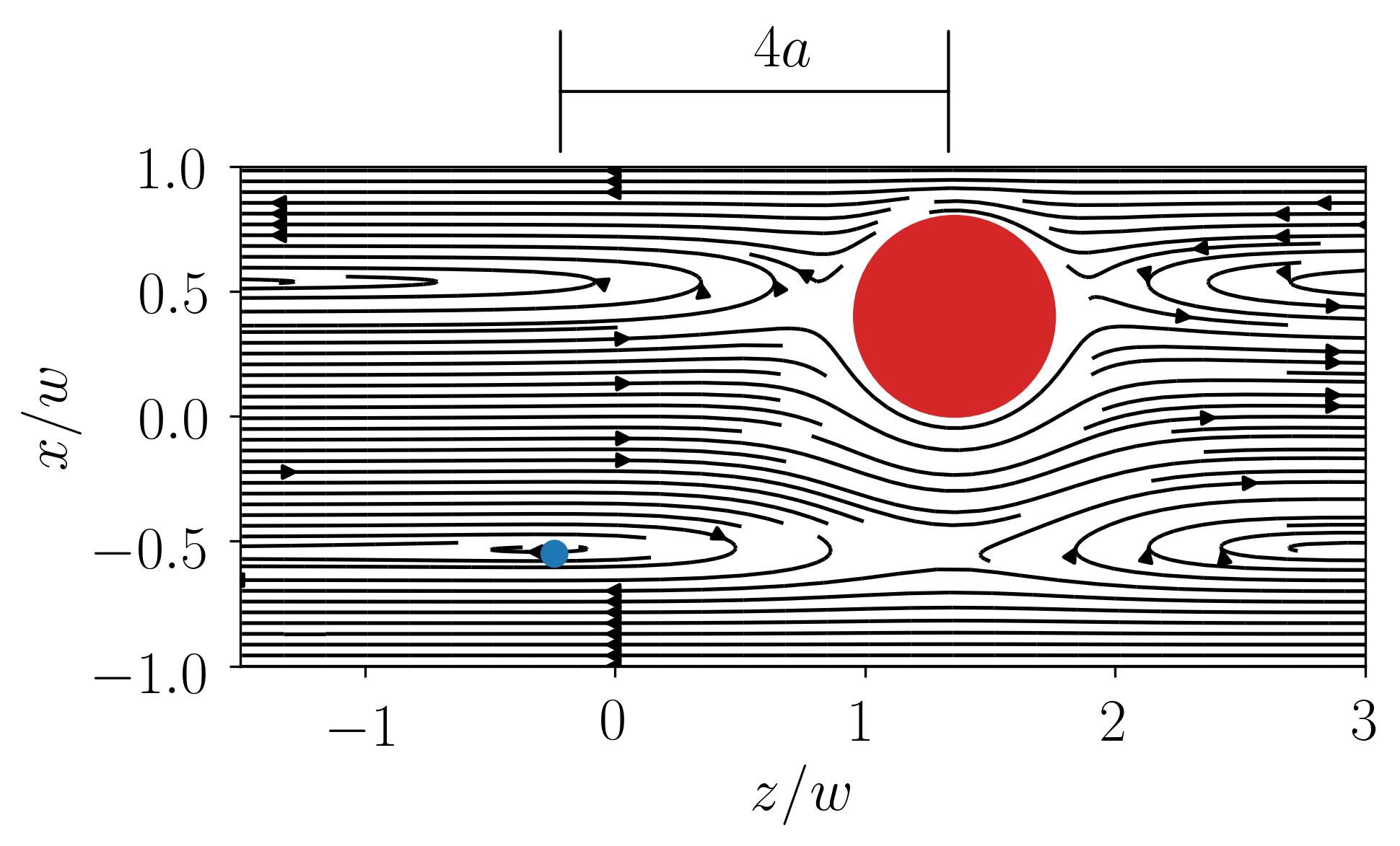}
	\caption{ \cs{Streamlines in the co-moving frame of a single particle 
	(red) at $\re=10$ show the formation of an eddy (blue)
	on the opposite side of the channel.}}
	\label{fig:streamline}
\end{figure}

\cs{In this article we explain the formation of stable particle pairs by the fact that the lift forces acting on them become zero and that both particles have the same distance from the channel center so that they drift with the same velocity. An alternative and intuitive explanation for the formation of cross-streamline pairs is given in Ref.\ \cite{humphry_axial_2010}. A particle creates a viscous disturbance flow, which contains eddies or vortices on the opposite side of the channel as indicated by the streamlines in the co-moving frame in \prettyref{fig:streamline}. The second particle then occupies the center of an eddy, where it does not move relative to the first particle. Since the viscous disturbance flow is independent of $\re$, the position of the eddy does not change with $\re$ in agreement with our argument in the previous paragraph. The advantage of the lift force profiles introduced in this article is that they not only describe equilibrium positions but that they also determine the full dynamics of a particle pair. We demonstrated this before when describing the oscillatory motion of the bound particle pair.}

\cs{In addition to the cross-streamline pairs \citet{hood_pairwise_2017} 
also formulated a theory that predicts
stable same-streamline pairs. However, in our analysis
particles moving on the same side of the channel never form bound states.
This result implies that two-particle interactions are not sufficient to explain particle lattices,
where all particle assemble on 
one side of the channel.}

\subsubsection{\cs{Unbound} trajectories}

\begin{figure}%
	\includegraphics[width=\linewidth]{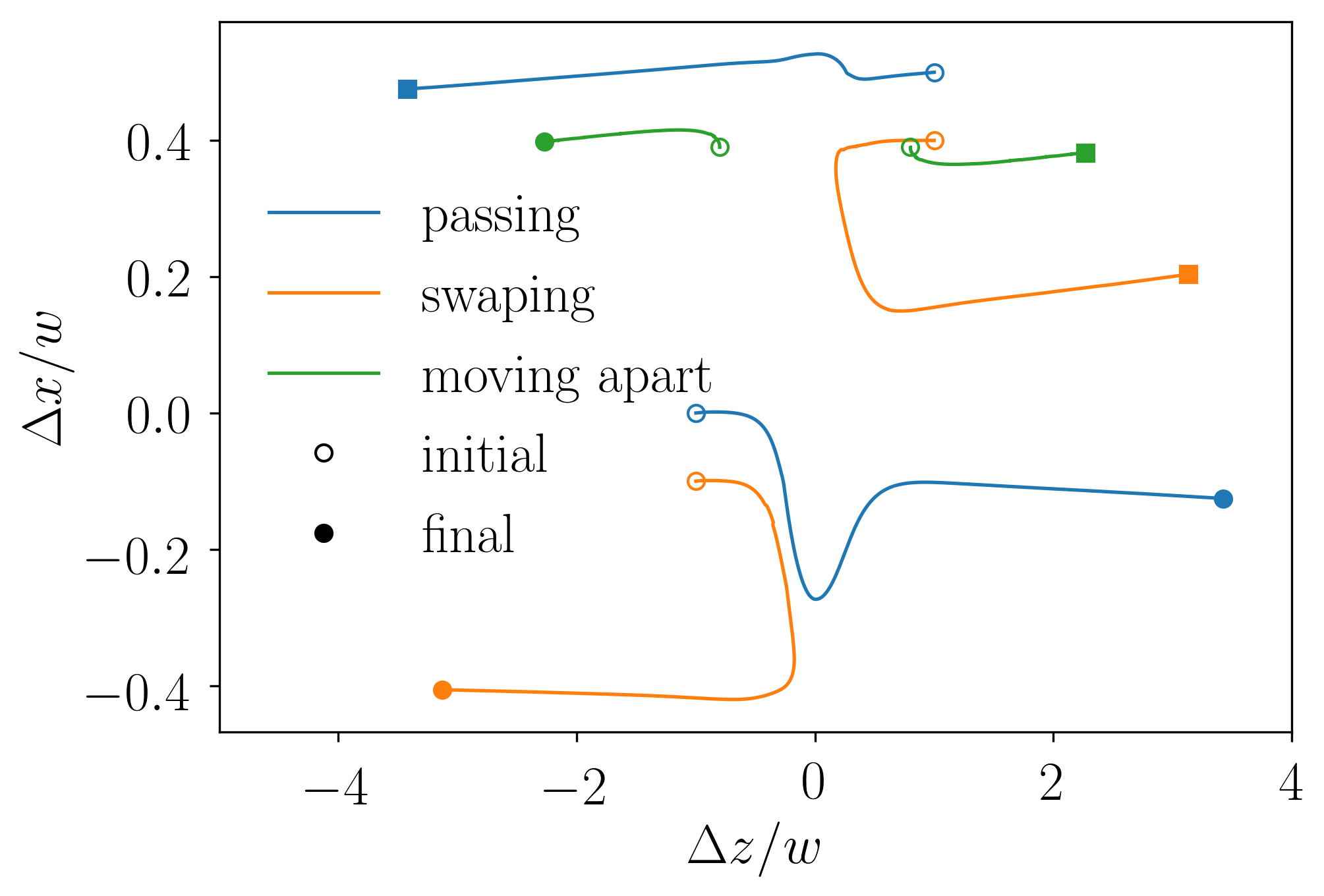}
	\caption{ %
		Three exemplary trajectories for \cs{unbound} 
		passing, swapping and moving-apart states drawn in the center-of-mass frame. Further parameters are $a/w=0.4$ and $\re=10$. Flow goes along the $z$ axis.}
	\label{fig:trajectory_unbound}
\end{figure}

We already introduced the \cs{unbound} trajectories, where no stable pair configuration 
exists \cs{(green,blue, and organge patches in  \prettyref{fig:trajectories})}.
In the moving-apart trajectories 
\cs{(green patch with $|x_{\mathrm{lead}}| < |x_{\mathrm{lag}}|$)}
the leading particle is faster than the lagging one.
\cs{The}
distance between both simply increases
\cs{and they independently migrate}
towards their equilibrium positions due to inertial focusing as given by the single-particle lift force profile.
Interestingly, even when the particles start 
on
their single-particle equilibrium positions
\cs{in the same half of the channel,}
they do not keep their distance fixed but
move on a moving-apart trajectory  (\cs{green line in} Fig.\ \ref{fig:trajectory_unbound}).
The reason is that these single-particle fixed points are not stable at 
a small particle distance, which also shows again the asymmetry in the force profiles.
The 
lift forces push
the leading particle closer to the center and the lagging particle closer to the walls. This 
causes a non-zero
relative velocity and 
the particles move apart ending at a larger axial distance.
\cs{Finally, in
the narrow green stripe with $x_\mathrm{lead}  > x_\mathrm{lag}$ (see \prettyref{fig:trajectories})
the particles initially approach each other. However, the lagging particle also drifts towards the wall so that it 
becomes slower than the leading particle and they both just move apart.}

Passing trajectories occur since the lagging particle is  closer to the channel center and therefore 
faster than the leading particle. So,
they change the order in axial direction (blue line in \prettyref{fig:trajectory_unbound}). During this overtaking the displacement of the two particles is asymmetric. The particle closer to the channel center is displaced much stronger and the offset is clearly visible 
after the passing event.
Then the axial distance increases and the particles
assume their single-particle positions due to inertial focusing. We note
that the particles do touch during the overtaking as we did not implement any lubrication approximation.
\cs{However, due to our event-based Euler step 
they do not overlap.}

When moving on swapping trajectories (orange line in \prettyref{fig:trajectory_unbound}), the faster lagging particle does no 
succeed to overtake the leading particle. Instead, 
\cs{the particles come close to each other and then swap the lateral position, which makes the leading particle the faster particle
so that they}
keep their axial order. 
When the particles move apart, they have interchanged their distances to the channel center. For example, for the orange trajectories in \prettyref{fig:trajectory_unbound} one finds
$x_\text{lead}^\text{after}\approx -x_\text{lag}^\text{before}$ and vice versa. \cs{Note similar trajectories in linear shear flow were called reversing trajectory \cite{kulkarni_pairsphere_2008}}

In general we see very similar types of trajectories also 
at low Reynolds numbers \cite{reddig_nonlinear_2013} indicating that 
they are governed by the viscous particle coupling and the Poiseuille flow profile. Inertial forces are responsible for focusing the particles on positions determined either by the two-particle or single-particle lift force profiles.
\cs{Although the two-particle trajectories studied in this article
are often unbound,
in preliminary results we find that they are the fundamental building blocks in the formation of multi-particle lattices.
Our goal is to explain the formation 
of particle lattices using also these trajectories in a 
future
work.}

\section{Conclusion}

Understanding 
pair interactions of two particles in inertial microfluidics is an important step for understanding collective dynamics such as the formation of particle trains. 

In this work we studied the lift force profiles and the trajectories of a pair of two solid particles driven by Poiseuille flow through a rectangular microchannel.
The lift force profiles of both particles are strongly influenced by their neighbors and depend on the particle distance along the channel axis. They clearly differ between the leading and lagging particles and the lift forces are stronger compared to a single particle.
The increased lift force should enhance 
particle focusing
by driving them faster towards their equilibrium positions. At close distance 
the lift force profiles 
differ
strongly from the profile of a single particle and do not allow for stable pair con\-fi\-gu\-rations.
However, when 
increasing the axial distance or 
the channel Reynolds number, the profiles appear 
similar in shape
but are shifted by constant forces.
Interestingly, at small  axial distances below $\Delta z/a = 4$ the strength of the lift forces scales with $\re$ indicating that hydrodynamic interactions between the particles are dominated by viscous forces, while for distances $\Delta z/a=10$ and larger scaling is quadratic in $\re$ showing the importance of inertial forces. In between, the scaling follows $\re^\alpha$ with the exponent varying smoothly from 1 to 2 while increasing $\Delta z$.
Finally, we presented the lift force profiles of leading and lagging particles in a two-dimensional representation
as a function of both lateral particle 
positions.

These two-dimensional plots determine the coupled dynamics and the trajectories of two floating particles. We identified four types of particle trajectories depending on the initial lateral position of the leading and lagging particles. Three of them are unbound, where
the particle distance ultimately increases until both particles reach their single-particle equilibrium positions. In the moving-apart trajectories
the leading particle is mostly faster than
the lagging particle and the pair drifts apart. If
the lagging particle is much faster, it overtakes and thereby changes axial order with the leading particle in what we call passing trajectories.
If the lagging particle is not much faster, it only approaches the leading particle but then they exchange their lateral positions and move apart again. Thus, they move on swapping trajectories. 
Finally, bound trajectories occur for $x_{\mathrm{lag}} \approx - x_{\mathrm{lead}}$, where axial distance and lateral positions of the particles perform damped oscillations while reaching
their equilibrium values.
As 
such a damping does not occur in Stokes flow, it is clearly an inertial effect. Consequently, the damping rate
 scales with the Reynolds number squared,
while the oscillation frequency increases linearly in $\re$.
Interestingly, for the specific particle radius studied here
all oscillating trajectories ultimately end at an axial distance of $\Delta z/a \approx 4.1$ independent of the initial 
conditions and the Reynolds number.

With our investigations we hope to shed further light on the col\-lective dynamics and ordering of particles flowing through micro\-fluidic channels at moderate Reynolds numbers. It should be useful in 
designing microfluidic crystal structures as well as developing and improving particle separation and sorting techniques in inertial microfluidics.

\begin{acknowledgments}
We thank C. Prohm for providing the source code and for initial work in the project. 
We 
acknowledge support from the Deutsche Forschungsgemeinschaft in the framework of the Collaborative 
Research Center SFB 910.
\end{acknowledgments}

\bibliography{literature}

\end{document}